\documentclass[fleqn]{annalen}
\usepackage{graphics,amssymb}
\begin{document}
\newcommand{\volume}{8}              
\newcommand{\xyear}{1999}            
\newcommand{\issue}{5}               
\newcommand{\recdate}{29 July 1999}  
\newcommand{\revdate}{dd.mm.yyyy}    
\newcommand{\revnum}{0}              
\newcommand{\accdate}{dd.mm.yyyy}    
\newcommand{\coeditor}{ue}           
\newcommand{\firstpage}{507}         
\newcommand{\lastpage}{510}          
\setcounter{page}{\firstpage}        
\newcommand{\keywords}{disorder, localization, metal-insulator
  transition, random hopping}
\newcommand{\PACS}{71.30.+h, 72.15.Rn, 72.80.Ng}
\newcommand{\shorttitle}{P. Cain et al., The 3D Anderson model of localization 
with random hopping}
\title{Phase diagram of the three-dimensional
 Anderson \\model of localization with random hopping }
\author{P. Cain, R. A. R\"omer, and M. Schreiber} 
\newcommand{\address}
  {Institut f{\"u}r Physik, Technische Universit{\"a}t, D-09107
  Chemnitz, Germany}
\newcommand{\email}{\tt cain@physik.tu-chemnitz.de} 
\maketitle
\begin{abstract}
  We examine the localization properties of the three-dimensional (3D) Anderson
  Hamiltonian with off-diagonal disorder using the transfer-matrix
  method (TMM) and finite-size scaling (FSS).  The nearest-neighbor
  hopping elements are chosen randomly according to $t_{ij} \in
  [c-1/2, c + 1/2]$.  We find that the off-diagonal
  disorder is not strong enough to localize all states in the spectrum
  in contradistinction to the usual case of diagonal disorder.  Thus
  for any off-diagonal disorder, there exist extended states and,
  consequently, the TMM converges very slowly.  From the TMM results
  we compute critical exponents of the metal-insulator transitions (MIT),
  the mobility edge $E_c$, and study the energy-disorder phase
  diagram.\hfill
\end{abstract}

\section{Introduction}
From the scaling hypothesis of localization \cite{Abrahams} it is well
known that in one (1D) and two dimensions (2D) almost all states of a
non-interacting disordered quantum system are localized \cite{KramKin93}.
Delocalized states occur only for special choices of physical
parameters. An example of such a model is provided by
the Anderson model of localization with random hopping \cite{Eilmes},
\begin{equation}
\label{eq-hamil}
  H = \sum_{i \neq j}^{N} t_{ij} |i \rangle \langle j| +
      \sum_i^N \epsilon_i |i \rangle \langle i| \quad .
\end{equation}
The hopping elements $t_{ij}$ between sites $i$ and $j$ are restricted
to nearest neighbors and chosen randomly from the
  interval $[ c - w/2, c + w/2 ]$.  Thus $c$ represents the center
and $w$ the width of the off-diagonal disorder distribution.  The
onsite potential energies $\epsilon_i$ are taken to be randomly
distributed in the interval $[-W/2,W/2]$. We set the energy scale by
keeping $w=1$ fixed and use periodic boundary conditions.

In this work we will concentrate on the properties of the random
hopping model in 3D. We study the density of states (DOS). Using TMM
\cite{KramKin83} together with FSS, we determine the mobility edge
$E_c$ of the MIT separating extended from localized states. We find
that the off-diagonal disorder is not strong enough to localize all
states in the spectrum in contradistinction to the usual case of
diagonal disorder.
\section{Density of States}
\begin{figure}[th]
\centerline{
\resizebox{0.5\textwidth}{0.43\textwidth}{\includegraphics{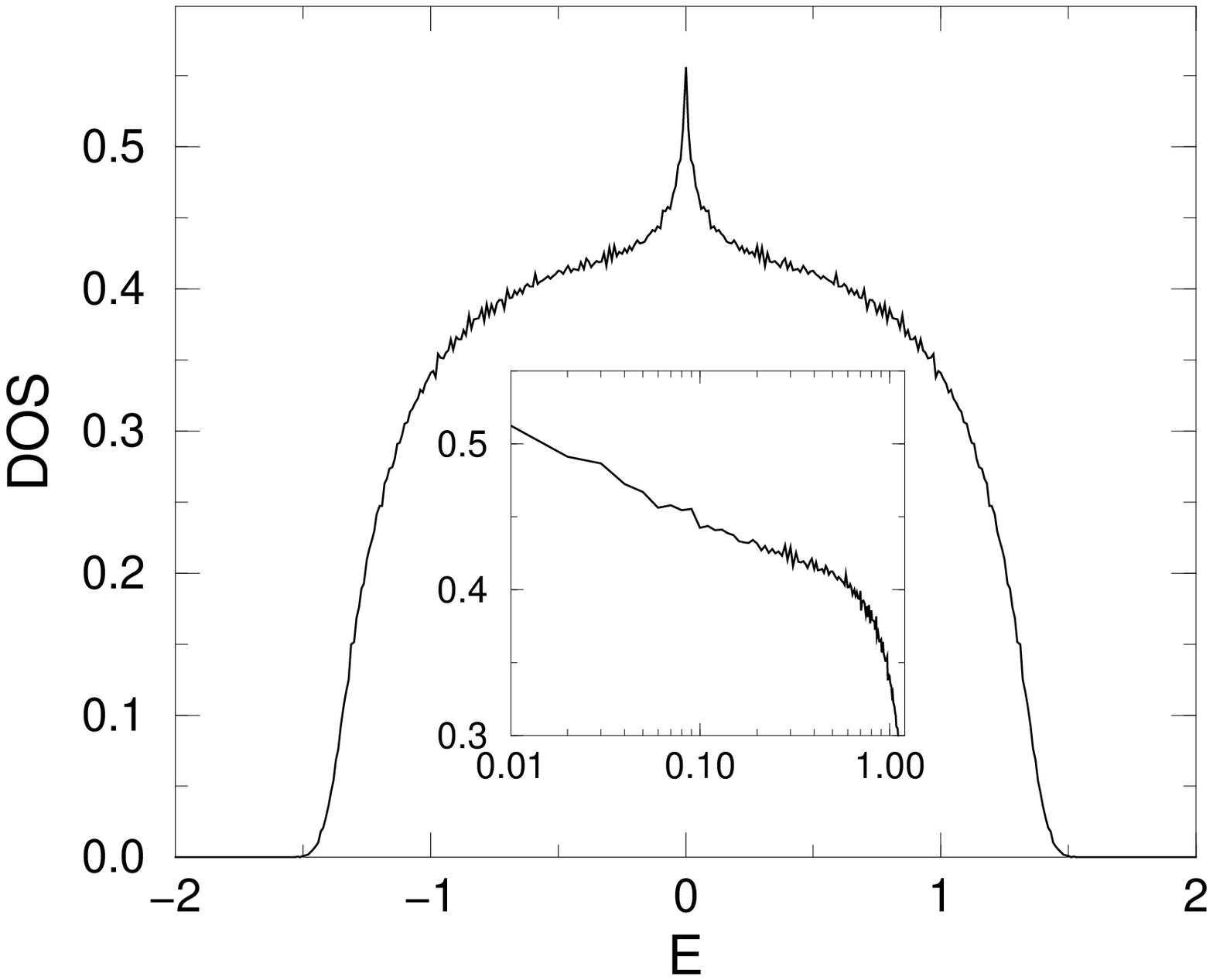}}
\resizebox{0.5\textwidth}{0.43\textwidth}{\includegraphics{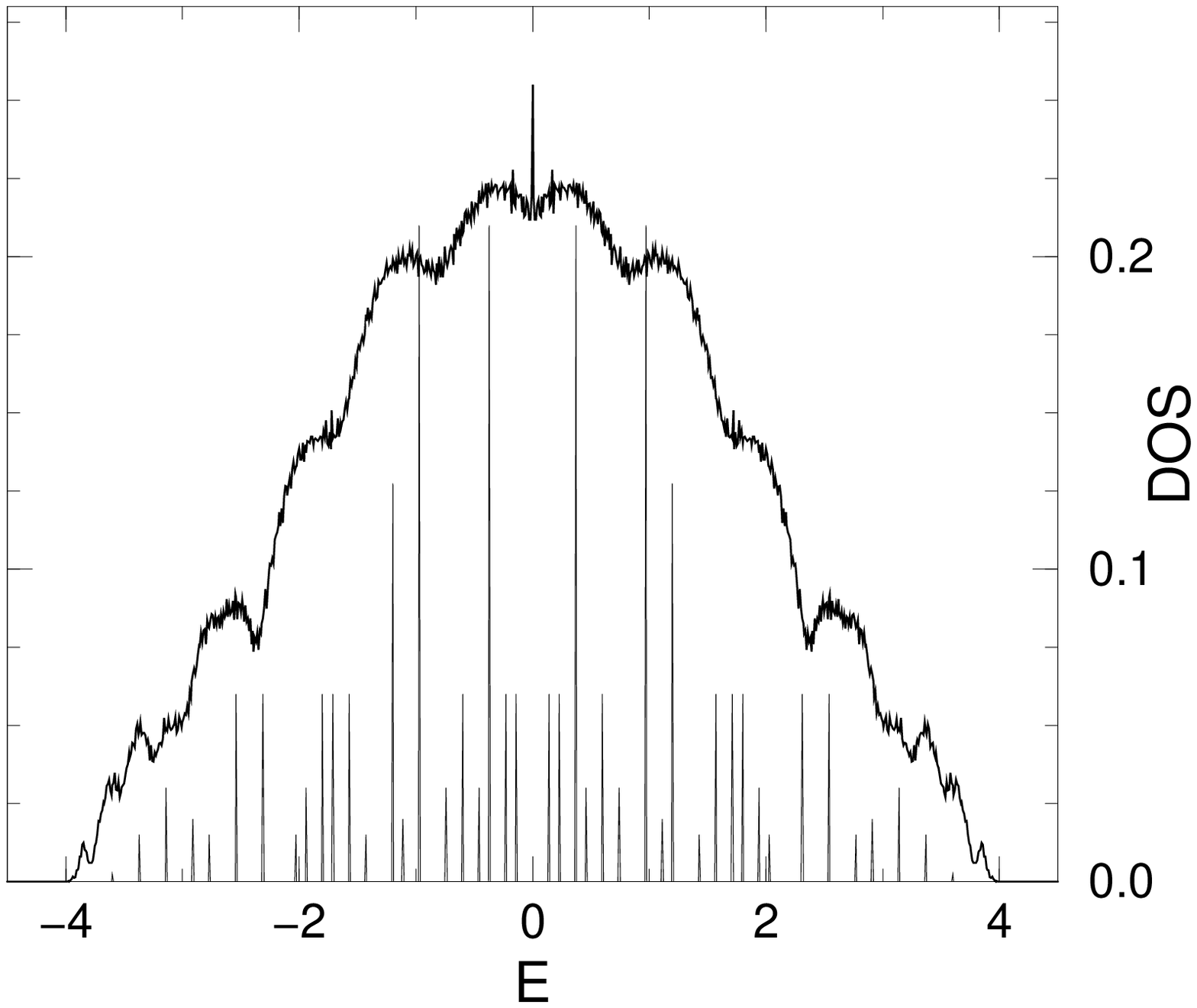}}}
  \caption{\label{fig-dos}\small 
    Left: DOS for off-diagonal disorder with system size $10^3$ and
    $c=0$; Inset: logarithmic behaviour of the DOS near the
    bandcenter.  Right: DOS for $c=0.6$; sharp lines correspond to the
    DOS of the ordered system at $w,W=0$, scaled by $1/40$ for
    clarity. The binwidth in all cases is $0.01$.}
\end{figure}

We have numerically computed the spectrum of the Hamiltonian
(\ref{eq-hamil}) for $1000$ samples, system size $N= M^3= 10^{3}$ and
various values of $c$.  In Fig.\ \ref{fig-dos} we show the resulting
DOS for strong off-diagonal disorder $c=0$ and moderate $c=0.6$. For
$c=0$ the peak at energy $E=0$ in the DOS of the finite system is well
pronounced. It corresponds to a logarithmic singularity in the
infinite system \cite{Dyson} as shown in Fig.\ \ref{fig-dos} (left,
inset).  For larger values of $c$, the off-diagonal disorder becomes
weaker and the DOS shows various sub-bands which reflect the spectral
structure of the completely ordered model for $w,W=0$. We
remark that in the corresponding 2D model the smallest localization
length and thus the strongest disorder was found at $c\approx 0.25$
\cite{Eilmes}.
\section{Determination of the mobility edge}
The application of the TMM for the present case of random hopping
relies on a reformulation of the Schr\"{o}dinger equation
\cite{KramKin93} as
\begin{eqnarray}
t^{||}_{jk n+1}\psi_{ jk n+1}
& = & (E-\epsilon_{jkn}) \psi_{jkn}- t^{||}_{jkn}\psi_{ jk n-1}\nonumber
 - t^{\perp}_{j+1kn} \psi_{j+1kn}
\\ & &-t^{\perp}_{jkn}\psi_{j-1kn} 
 - t^{\perp^\prime }_{jk+1n}\psi_{jk+1n}-t^{\perp^\prime }_{jkn}\psi_{jk-1n}
\label{eq-tmm}
\end{eqnarray}
with $\psi_{jkn}$ corresponding to the wave function at site
$(j,k,n)$, $t^{\perp}_{jkn}$ and $t^{\perp^\prime }_{jkn}$ representing the hopping elements from
site $(j-1,k,n)$ or $(j,k-1,n)$ respectively to site $(j,k,n)$ and ${t^{||}_{jkn}}$ the hopping
element from $(j,k,n-1)$ to $(j,k,n)$. This equation may be written in
matrix from as
\begin{eqnarray}
  \left( \begin{array}{l} \psi_{n+1} \\ \psi_{n} \end{array} \right) 
=
T_n
\left( \begin{array}{l} \psi_{n} \\ \psi_{n-1} \end{array} \right)
,\quad
T_n 
=
\left(
 \begin{array}{cc}
   [{t^{||}_{n+1}}]^{-1}(E - H_{\perp})  
& -[{t^{||}_{n+1}}]^{-1}
{t^{||}_{n}} \\ 
   1              
&  0
 \end{array}
\right)
\end{eqnarray}
with
$\psi_n=(\psi_{11n},...,\psi_{1Mn},\psi_{21n},...,\psi_{MMn})$
denoting the wave function in the $n\rm th$ slice, $H_\perp$ the
Hamiltonian in the $n\rm th$ slice, and the matrix ${t^{||}_{n}}$
connecting the $(n-1)\rm th$ with the $n\rm th$ slice.
As usual, the localization length $\lambda = 1/\gamma_{\rm
  min}$ is computed from the smallest Lyapunov exponent $\gamma_{\rm
  min}$ of the $2M^2$ eigenvalues $\exp (\pm \gamma_{i})$ of $\Gamma=
\lim_{K\rightarrow\infty}(\tau_{K}^\dagger \tau_{K})^{1/2K}$. Here,
$\tau_{K} = T_{K} T_{K-1} \cdots T_{2} T_{1}$.  Assuming the validity
of one-pa\-rameter scaling close to the MIT, we expect the reduced
localization lengths $\lambda(M)/M$ to scale onto a scaling curve
$\lambda(M)/M = f(\xi/M)$ with scaling parameter $\xi$.

\begin{figure}
\centerline{
\resizebox{0.5\textwidth}{0.45\textwidth}{\includegraphics{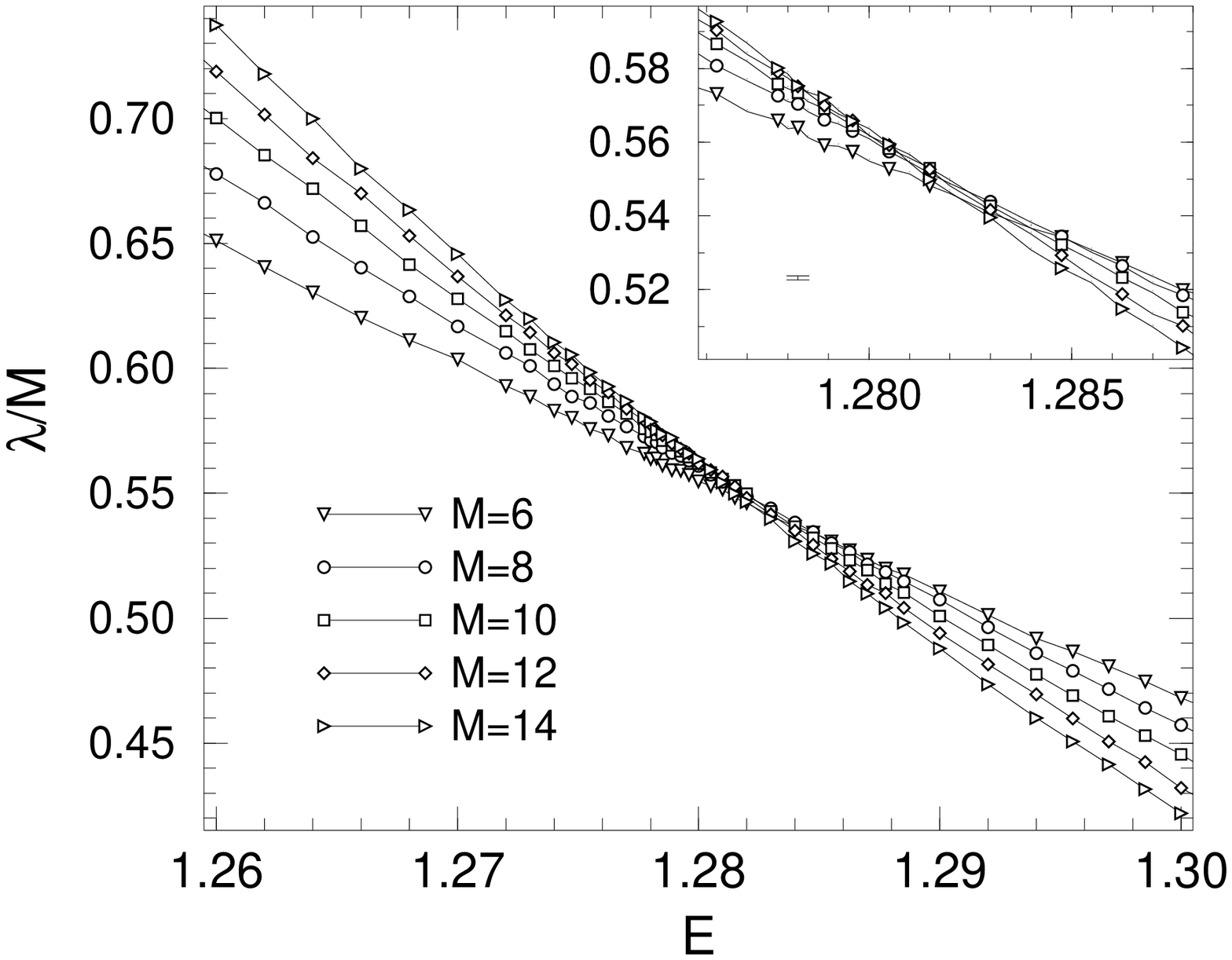}}
\resizebox{0.5\textwidth}{0.45\textwidth}{\includegraphics{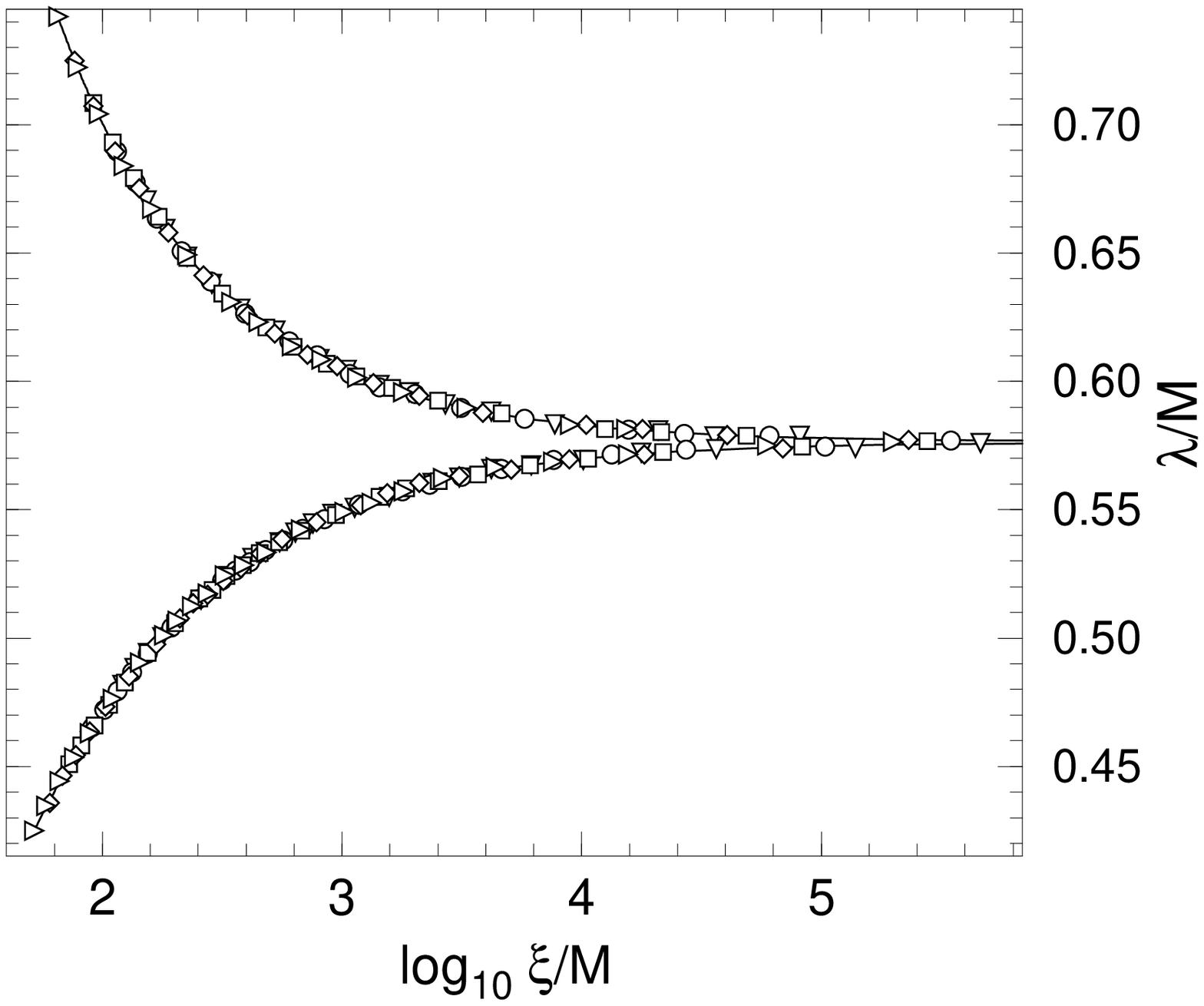}}}
  \caption{\label{fig-fss}\small 
    Left: Reduced localization length $\lambda/M$ as function of
    energy $E$ at $c=0$ and $W=0$ as obtained by TMM with $0.1\%$
    accuracy.  Note the MIT at $E_c\approx 1.28$. Inset: Enlarged
    region close to the MIT, only every second data point indicated by
    a symbol, the error bar shown separately. Right: FSS curve for the
    TMM data from the left panel.  }
\end{figure}

In Fig.\ \ref{fig-fss} we show the results up to $M=14$ at $c=0$ and $W=0$. The
crossing between extended states at $E \lesssim 1.28$ and localized states
at $E \gtrsim 1.28$ is clearly visible. The mobility edge is thus
$E_c\approx 1.28$. The inset zooms into the MIT region, where one can
see an $M$-dependent shift of the crossing point towards smaller $E$
for increasing $M$. Due to this finite size effect, which is supposed
to vanish for $M$ large enough, we take non-linear and non-universal
corrections to FSS \cite{Slevin} into account.
The FSS curve constructed from these data in
Fig.\ \ref{fig-fss} shows two branches
similarly to the FSS curves in the Anderson model of
localization with pure diagonal disorder \cite{Schreiber}. However,
in contradistinction to the model with diagonal disorder, we find
that the off-diagonal disorder at $c=0$ is not strong enough to
localize all the states.

\begin{figure}[th]
\centerline{\resizebox{7.1cm}{!}{\includegraphics{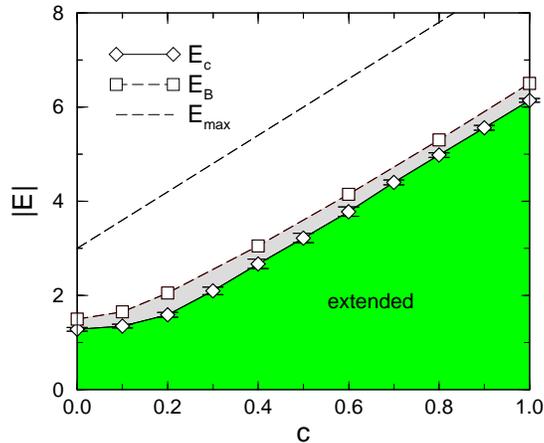}}}
  \caption{\label{fig-mobedge}\small 
    Phase diagram of the Anderson model of localization with
    off-diagonal disorder $c$ as obtained from TMM with $1\%$
    accuracy.  The values of $E_c$ are indicated by diamonds. Squares
    represent the numerically determined upper band edge $E_B$ for a
    system with $10^{3}$ sites averaged over $1000$ samples. The
    straight dashed line is the band edge $E_{\rm max}$ for the
    infinite system corresponding to a constant $t=c+1/2$. The
    difference between $E_B$ and $E_{\rm max}$ occurs due to
    exponential tails of the DOS which cannot be resolved by our
    calculations.  States with $|E|<E_c$ are extended (dark gray),
    whereas states with $E_c < |E| \leq E_B$ are localized (light
    gray). }
\end{figure}

Analysing data for larger values of $c$, we find increasing
localization lengths. Consequently the mobility edge
increases with increasing $c$. In Fig.\ \ref{fig-mobedge} we show
the results up to $c=1$.
For larger $c$ the transition observed for $c=0$ in Fig.\ 
\ref{fig-fss} becomes less easy to detect and we obtain reasonably
accurate crossings for large $M$ only.  The crossing becomes sharp
again when varying $W$ instead of $E$.  This allows us to determine
the critical exponents $\nu_E$ and $\nu_W$ of the scaling parameter
$\xi \propto |E-E_c|^{\nu_E}$ and $\xi \propto |W-W_c|^{\nu_W}$.  In
agreement with recent results in the Anderson
model with pure diagonal disorder \cite{Slevin} we find $\nu_E = 1.61\pm0.07$ and $\nu_W=1.54\pm
0.03$.
\section{Conclusions}
We have investigated the DOS and the localization properties of the
Anderson model of localization with off-diagonal disorder. Our results
show that in 3D a finite part of the spectrum close to the band center
remains extended even for the strongest off-diagonal disorder $c=0$,
whereas in 1D and 2D the off-diagonal disorder has non-localized
states only for $E=0$. FSS is possible and we determine the mobility
edge separating extended and localized states in the off-diagonal
disorder vs.\ energy phase diagram. The obtained critical exponents of
the MITs are in agreement with prior results for the model with only 
diagonal disorder \cite{Slevin,MacKinnon94} and
convincingly support the universality.

%
%
%
%
%
%
%
%
%
%
%
%

\end{document}